\documentclass[aps,prl,epsfig,graphics,twocolumn,floatfix,mathbbm,a4paper]{revtex4}

\usepackage{amsmath,amsfonts,amssymb,graphics,graphicx,epsfig,color,times,bbm}

\begin{document}
\bibliographystyle{apsrev}

\newcommand{\R}{\mathbbm{R}}
\newcommand{\rr}{\mathbbm{R}}
\newcommand{\E}{{\cal E}}
\newcommand{\V}{{\cal V}}
\newcommand{\cc}{{\cal{C}}}
\newcommand{\ii}{\mathbbm{1}}

\newcommand{\1}{\mathbbm{1}}
\newcommand{\F}{\mathbbm{F}}
\newcommand{\h}{\frak{H}}
\newcommand{\tr}[1]{{\rm tr}\left[#1\right]}
\newcommand{\gr}[1]{\boldsymbol{#1}}
\newcommand{\be}{\begin{equation}}
\newcommand{\ee}{\end{equation}}
\newcommand{\bea}{\begin{eqnarray}}
\newcommand{\eea}{\end{eqnarray}}
\newcommand{\ket}[1]{|#1\rangle}
\newcommand{\bra}[1]{\langle#1|}
\newcommand{\avr}[1]{\langle#1\rangle}
\newcommand{\D}{{\cal D}}
\newcommand{\eq}[1]{Eq.~(\ref{#1})}
\newcommand{\ineq}[1]{Ineq.~(\ref{#1})}
\newcommand{\sirsection}[1]{\section{\large \sf \textbf{#1}}}
\newcommand{\sirsubsection}[1]{\subsection{\normalsize \sf \textbf{#1}}}
\newcommand{\ack}{\subsection*{\normalsize \sf \textbf{Acknowledgements}}}
\newcommand{\front}[5]{\title{\sf \textbf{\Large #1}}
\author{#2 \vspace*{.4cm}\\
\footnotesize #3}
\date{\footnotesize \sf \begin{quote}
\hspace*{.2cm}#4 \end{quote} #5} \maketitle}
\newcommand{\eg}{\emph{e.g.}~}

\newcommand{\proofend}{\hfill\fbox\\\medskip }

%---------------------------------------------------------------------------

\newtheorem{theorem}{Theorem}
\newtheorem{proposition}{Proposition}

\newtheorem{lemma}{Lemma}

\newtheorem{definition}{Definition}
\newtheorem{corollary}{Corollary}

\newcommand{\proof}[1]{{\it Proof.} #1 $\proofend$}

\title{Assessing dimensions from evolution}

\author{Michael M. Wolf$^1$, David Perez-Garcia$^2$}
\address{$^1$ Niels Bohr Institute, 2100 Copenhagen, Denmark\\
$^2$ Dpt. An\'alisis Matem\'atico,
Universitad Complutense de Madrid, 28040 Madrid, Spain}
\date{\today}

\begin{abstract}Using tools from classical signal processing, we show how to determine the dimensionality of a quantum system as well as the effective size of the environment's memory from observable dynamics in a model-independent way. We discuss the dependence on the number of conserved quantities, the relation to  ergodicity and prove a converse showing that a Hilbert space of dimension $D+2$ is sufficient to describe every bounded sequence of measurements originating from any $D$-dimensional linear equations of motion. This is in sharp contrast to classical stochastic processes which are subject to more severe restrictions: a simple spectral analysis shows that the gap between the required dimensionality of a quantum and a classical description of an observed evolution can be arbitrary large.
\end{abstract}

\maketitle

%%%%%%%%%%%%%%%%%%%%%%%%%%%%%%%%%%%%%%%%%%%%%%%%%%%%%%%%%%%%%%%%%%%%%%

%\begin{multicols}{2}

In Quantum Information Science the dimension of the accessible Hilbert space has the character of a resource---larger dimensions mean potentially more powerful protocols. Various implementations deal with huge Hilbert spaces corresponding to ensembles of atoms~\cite{HSP} or molecules~\cite{molecules}, or continuous degrees of freedom leading to an infinite dimensional space in the first place.
But how many degrees of freedom are effectively used? Can we assess the dimension of the underlying system from observable data without assuming a detailed model description beforehand? These questions, partially motivated by the need of questioning and pinpointing the assumptions of security proofs in quantum key distribution~\cite{QKD}, were recently addressed in the context of `non-local' quantum correlations.

In \cite{Marius} it was shown that a tripartite system can only yield violations of certain Bell inequalities of the order of $\sqrt{d}$ if each subsystem has dimension at least $d$, and in \cite{Brunner,recent}  the dimension dependence of correlations has been investigated in detail for bipartite systems. As these approaches are based on static correlations between several parts, the question has been raised~\cite{Brunner} whether and how the dimension of a single system can be observed.

Similarly, one might want to have a preferably model-independent way of assessing the effective dimensionality of the systems environment (quantifying non-Markovianity \cite{snapshot}) or the number of preserved, `noiseless' degrees of freedom.

The present work addresses these question from a dynamical point of view. Given a discrete time evolution of an expectation value, we ask what can be inferred about the effective dimension of the systems Hilbert space or the environments memory. We thereby focus on using as little a priori information as possible. When addressing the systems dimensionality, the only assumptions are that the evolution is homogenous in time and Markovian in the sense that it is performed on time scales large compared to the relevant relaxation times of the systems environment. If the latter is not fulfilled we will see the environmental memory degrees of freedom in our dimension count.

For this analysis two standard tools from classical signal processing~\cite{signalprocessing} can be employed: delayed embeddings and analysis in the frequency domain.
These will allow us not only to tackle the above question but also to address the converse: which Hilbert space dimension is sufficient for a given sequence of measurements? and to compare the efficiencies of quantum versus classical descriptions of a given evolution. While every sequence produced by quantum mechanical evolution can in principle be described by a classical stochastic process, we will easily see that quantum mechanics can be arbitrarily more efficient in terms of the required number of discrete degrees of freedom.

Not surprisingly we will see a connection between the effective dimension and the number of conservation laws and get a glimpse on more (in particular spectral) information which can be obtained from the evolution.

\section{Preliminaries}
Our interest lies in the discrete time evolution of expectation values of the form \be\label{eq:series} \avr{A(t)}=\tr{AT^t(\rho)},\ee where $\rho$ is a density matrix acting on a $d$-dimensional Hilbert space, $T$ is quantum channel, i.e., a trace-preserving completely positive linear map with equal input and output space and $A=A^\dagger$ an observable. For simplicity we will in the first part assume that (\ref{eq:series}) is a half-infinite sequence, i.e., $t\in\mathbb{N}_0=\{0,1,\ldots\}$; the extension to finite sequences and noisy data will then be discussed at the end.

Note that the description in (\ref{eq:series}) assumes homogeneity in time and \emph{Markovness} in the sense that future evolution only depends on the state at present and not on its history. This means that given the l.h.s. of (\ref{eq:series}) the Hilbert space underlying the description of the r.h.s. has to contain all effectively relevant degrees of freedom. Hence, if a system of dimension $d_S$ undergoes a non-Markovian evolution due to $d_E$ memory degrees of freedom in the environment, then $d=d_S+d_E$.

It will sometimes be advantageous to consider $\rho$ and $A$ as elements of a $d^2$-dimensional Hilbert space $\cal H$ equipped with the \emph{Hilbert-Schmidt} scalar product $\langle A|\rho\rangle:=\tr{A^\dagger \rho} $. As a linear map $T$ has a matrix representation on $\cal H$ which we will denote by $\hat{T}$. Using matrix units as a basis of $\cal H$ we can write $\hat{T}=\sum_k K_k\otimes \overline{K}_k$, where $\{K_k\}$ is a set of \emph{Kraus operators} of $T(\cdot)=\sum_k K_k\cdot K_k^\dagger$. While $T$ refers to time evolution in the Schr\"odinger picture, we will denote by $T^*$ the respective map in the Heisenberg picture so that $\tr{A T(\rho)}=\tr{T^*(A)\rho}$.

We will denote by \be{\cal C}:=\big\{H=H^\dagger\big|\avr{H(t)}\; \text{independent of } t\in\mathbb{N}_0\big\}\ee
the space of \emph{conserved quantities} which obviously includes all $H=T^*(H)$ and in particular $\1\in\cal C$ as the evolution is trace preserving. Note that $\cal C$ depends on $T$ and $\rho$.

A quantum channel $T$ will be called \emph{ergodic} w.r.t. a state $\rho$ (an observable $A$) if the orbit generated by  $T^{t}$ ($T^{*t}$) spans the entire space of $d\times d$ matrices.

\section{Assessing the dimension}
The central tool in this section is the space $\V$ spanned by all \emph{delayed vectors} of the form \be v_\tau=\big(\avr{A(\tau)},\avr{A(\tau+1)},\ldots\big),\quad \tau\in\mathbb{N}_0.\label{eq:vs}\ee The employed approach, often called \emph{method of delays}, is particularly widespread in the analysis of chaotic dynamics \cite{Takens} and it provides the following simple and tight relation:

\begin{proposition}[A bound on the dimensionality]\label{prop:dim} Consider the space $\V=\mbox{span}\{v_\tau\}_{\tau\in\mathbb{N}_0}$ spanned by the delayed vectors obtained from a sequence of the form (\ref{eq:series}).  Then
\be \dim{\cal C}+\dim \V\leq d^2+1, \label{eq:mainIneq} \ee
where $d$ is the dimension of the underlying Hilbert space and $\dim\cal C$ the number of linearly independent conserved quantities. Equality holds in (\ref{eq:mainIneq}) if $T$ is ergodic w.r.t. $A$ and  $\dim \V=d^2$ iff it is ergodic w.r.t. $A$ and $\rho$.
\end{proposition}
\proof{ Consider a basis $\{H_i\}$, $i=1,\ldots,D:=\dim\cal C$ of the space $\cal C$ of conserved quantities. Then $\{H_i-\tr{\rho H_i}\1\}$ span a  $D-1$ dimensional subspace of $\cal H$ which is orthogonal to the space spanned by $\hat{T}^t|\rho\rangle$, $t\in\mathbb{N}_0$. Hence the latter space, denote it by $\h$, has dimension $d^2-D+1$ and \be
\avr{A(t)}=\langle A_{\h}| \hat{T}^t_{\h}|\rho_{\h}\rangle,\ee
where the index $_{\h}$ refers to the restriction onto $\h$. Since the minimal polynomial~\cite{Gantmacher} of $\hat{T}_\h$ has degree at most $\dim\h$ there are complex coefficients $c_j$ such that \be \hat{T}_\h^{\dim\h}=\sum_{j=0}^{d^2-D} c_j \hat{T}_\h^j. \ee Recalling that $(v_\tau)_k=\avr{A(\tau+k-1)}$ this implies that there are at most $\dim \h$ linearly independent vectors in $\V$ since \be (v_{d^2-D+x})_k=\sum_{j=0}^{d^2-D} c_j\; (v_j)_{x+k-1}\quad\forall x,k\in\mathbb{N}.\ee
Let us now assume that $T$ is ergodic w.r.t. $A$, which means that $\{\langle A|T^t\}_{t\in\mathbb{N}_0}$ spans $\cal H$. Suppose that (\ref{eq:mainIneq}) would not be an equality. Then $\sum_{j=0}^{d^2-D} c_j v_j=0$ for some $c$ which implies, by ergodicity, that $\sum_{j=0}^{d^2-D} c_j \hat{T}^{j+n}|\rho\rangle=0$ for all $n\in\mathbb{N}_0$. Therefore $\dim\h\leq d^2-D$ so that for the orthogonal complement $\dim\h^\perp\geq D$. However, this contradicts the inequality $\dim\h^\perp\leq D-1$ which comes from the fact that every element in $\h^\perp$ is in $\cal C$ and in addition $\1\in{\cal C}\setminus\h^\perp$. So, ergodicity w.r.t. $A$ implies equality in (\ref{eq:mainIneq}).

If $T$ is in addition ergodic w.r.t. $\rho$ then $\h=\cal H$ implies $D=1$ so that indeed $\dim \V=d^2$.
 We finally prove the converse again by contradiction:  assume linear dependence of the form $\sum_{j=0}^{d^2-1}c_j v_j=0$. Then for all $a,b\in\mathbb{N}_0$ we have  $\langle A|\hat{T}^a \big(\sum_j c_j \hat{T}^j\big) \hat{T}^b|\rho\rangle=0$ so that, due to ergodicity, $\hat{T}$ would have a minimal polynomial of degree less than $d^2$. However, this would again imply the existence of a proper subspace $\h$ contradicting the assumption $\dim \V=d^2$. To see this recall that a minimal polynomial of smaller degree requires an eigenvalue $\lambda$ with geometric multiplicity larger than one~\cite{Gantmacher}. Denoting by $|\phi\rangle$ a linear combination of the corresponding left eigenvectors, we get $\langle \phi|\hat{T}^t|\rho\rangle=\lambda^t\langle \phi|\rho\rangle$ such that there is always a $\phi\perp\h$.
}

\emph{Some remarks.} Depending on the assumptions we may use  the above result for different purposes: (i) assuming Markovianity it provides a lower bound for $d$, (ii) assuming we know $d$ in addition it yields an upper bound on $\dim\cal C$ and (iii) if we only know the dimension of the system $d_S=d-d_E$ it gives a lower bound on the number $d_E$ of effective memory degrees of freedom in the environment. In fact, if $\langle A(t)\rangle$ exhibits algebraic decay then, as one would expect, Eq.(\ref{eq:mainIneq}) leads to $d=\infty$.

Prop.~\ref{prop:dim} is easily generalized to the case where, instead of a single expectation value, we observe a set of observables $\{A_\alpha\}$ or, equivalently, take higher moments of the observable into account (i.e., $A_\alpha=A^\alpha$). Then the delayed vectors have to be replaced by `delayed matrices' so that $\langle A_\alpha(\tau)\rangle$ are the entries of the first column of the matrix $v_\tau$. Eq.(\ref{eq:mainIneq}) is then obtained in just the same way where $\dim\cal V$ is now the number if linearly independent matrices.

\section{Quantum evolution for given sequences}
The previous section provided a lower bound on the dimension of the Hilbert space in terms of the dimension of the space $\V$ of delayed vectors which are in turn solely based on $\langle A(t)\rangle$. Clearly, there cannot be an upper bound to the `true' dimension since there might always be redundant degrees of freedom. Nevertheless, we  can provide a converse to the above observation in the sense that there always exists a quantum representation in a Hilbert space of dimension not much larger than $\dim \V$. Remarkably, such a converse does not exist for classical evolutions (see subsequent section).
\begin{proposition}[Quantum representation]\ \\
Given any bounded sequence $\avr{A(t)}\in\mathbb{R},\;t\in\mathbb{N}_0$ there is always a quantum state $\rho$, a quantum channel $T$ and a Hermitian observable $A$ acting on a Hilbert space of dimension $\dim \V+2$ such that (\ref{eq:series}) holds.
\end{proposition}
\proof{We begin with the fact proven in Lemma \ref{lem:linear} in the appendix: there is always a contractive matrix $M$ and vectors $R,L$ of dimension $\dim\V$ such that $\avr{A(t)}=\langle L|M^t|R\rangle$. We will proceed in two steps: first establish complete positivity by adding one degree of freedom and then impose the trace preserving property by adding another degree. Define a square `Kraus-operator' $C=1\oplus M$ of dimension $\dim\V+1$ and, referring to the same block structure ($\mathbb{C}\oplus\mathbb{C}^{\dim\V}$), a vector $|\Psi\rangle=|0\oplus R\rangle+|1\oplus 0\rangle$ and an `observable' $B=(|1\oplus 0\rangle\langle0\oplus L|+h.c.)/2$. Then, using that $\avr{A(t)}\in\mathbb{R}$ we have $\langle L|M^t|R\rangle=\tr{B\; C^t|\Psi\rangle\langle\Psi|C^{\dagger t}}$. In order to make this trace preserving we embed it again, now referring to the block structure $\mathbb{C}\oplus\mathbb{C}^{\dim\V+1}$. With $K:=0\oplus C, A:=(0\oplus B)||\Psi||^2, \rho:=(0\oplus|\Psi\rangle\langle\Psi|)/||\Psi||^2$ we obtain that \be
T(\rho)=K\rho K^\dagger+ |1\oplus 0\rangle\langle 1\oplus 0|\tr{(\1-K^\dagger K)\rho}
\ee indeed satisfies (\ref{eq:series}) for the chosen $\rho$ and $A$, and since $||M||_\infty\leq 1$ implies $K^\dagger K\leq \1$, $T$ is a valid quantum channel.
}
\begin{figure}[ttt]
\begin{center}
%\psfrag{title}{}
%\psfrag{xlabel}{\Large $\alpha$}
%\psfrag{ylabel}{\Large Benchmarks}
%\psfrag{legend1}{\large $\overline{F}$}
%\psfrag{legend2}{\large $\overline{F}_{infinite}$}
%\psfrag{legend3}{\large $\overline{F}_{finite}$}
\includegraphics[width=8cm]{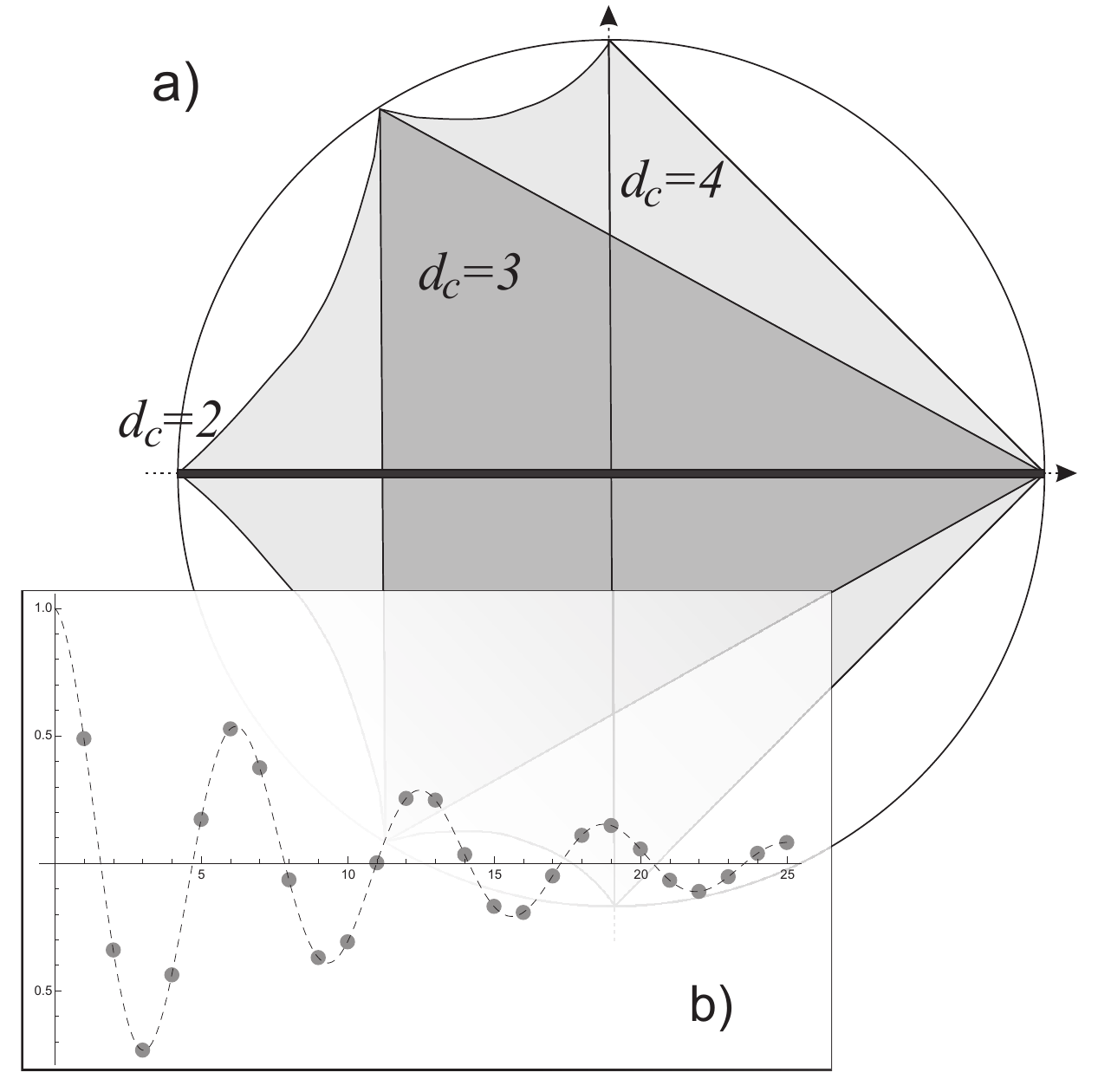}
%\epsfig{file=positivespectrum2in1F.eps,angle=0,width=0.92\linewidth}
%\vspace*{-0.4cm}
\end{center}
\caption{a) Eigenvalues of the evolution operator lie on the unit disc and can be obtained from the observable time dependence of any expectation value.  For a classical description they have to be located in a region depending on the number $d_c$ of degrees of freedom. For $d_c=2$ this is the real line, for $d_c=3,4$ the dark and light grey regions become additionally  accessible. b) A simple damped oscillation leads to eigenvalues $e^{-\gamma\pm i\omega}$, implying that a quantum mechanical description can access the entire unit disc even for $d=2$. }
\label{Fig:spectrum}
\end{figure}

We do not know whether there are more efficient constructions.
Clearly, from Prop.\ref{prop:dim} we know that one needs at least
$d\geq \sqrt{\dim\V}$, but this will most likely not be achievable
in general.

\section{Spectral analysis and separation from classical descriptions}

So far we investigated the dimensionality of the system based on the sequence $\avr{A(t)}$. The information obtained can be refined when going to the \emph{frequency domain} by considering the function ${\cal L}:\mathbb{C}\rightarrow\mathbb{C}$ (the `$z$-transform'~\cite{signalprocessing}) defined by the series
\be {\cal L}(z):=\frac{1}{z}\sum_{t\in\mathbb{N}_0}\frac{\avr{A(t)}}{z^t}.\label{eq:L}\ee
This converges outside the unit circle and can be define inside by analytic continuation. In this way we obtain for a sequence of the form (\ref{eq:series})
\be {\cal L}(z)=\tr{A\big(z\; {\rm id}-T\big)^{-1}(\rho)},\ee
so that poles of $\cal L$ correspond to eigenvalues of $T$. Note that the latter lie always inside (or on) the unit circle, there is an eigenvalue 1, and complex eigenvalues come in conjugate pairs. While there are restrictions \cite{dividing} for instance for the determinant, i.e., the product of eigenvalues, quantum mechanics does not impose any further constraint on the location of eigenvalues: any point on the unit disc is possible even for $d=2$. The simplest example for this is a damped Rabi oscillation leading to $\avr{A(t)}=e^{-\gamma t}\cos{\omega t}$ with poles of $\cal L$ at $e^{\pm i\omega-\gamma}$.

It is instructive to compare this with a potential classical description of the sequence $\avr{A(t)}$. So assume there are $d_c$ states to each of which we assign an initial probability $p_k$, $k=1,\ldots, d_c$. The evolution of these probabilities for a single time-step is governed by a stochastic matrix $S$ and in the end  a measurement outcome $a_k\in\mathbb{R}$ is assigned to the $k$'th state. In this way we arrive at \be \avr{A(t)}=\langle a|S^t|p\rangle\label{eq:classical}.\ee

Yet, the poles of ${\cal L}$ correspond to eigenvalues of $S$, which share the basic properties mentioned above. However, the classical description imposes additional constraints on the location of the eigenvalues depending on the dimension $d_c$. In particular, they have to be located inside the convex hull of all roots of unity up to order $d_c$. That is, the unit disc will not be entirely covered for any finite $d_c$. A more complete characterization of the location of eigenvalues is given in \cite{Karpelevich,DD} and shown in Fig.\ref{Fig:spectrum} for $d_c=2,3,4$.

A simple consequence of this analysis is that in terms of the required degrees of freedom a quantum mechanical description of a sequence $\avr{A(t)}$ can be far more efficient than a classical one. In the above discussion the separation between a quantum and a classical description comes from the simple fact that oscillations are easier to describe in terms of probability amplitudes than by using probabilities. Certainly more sophisticated separations can be found, however, a complete determination of $d$ and $d_c$ from a given sequence $\avr{A(t)}$ seems to be a daunting task (despite considerable results on the classical side, cf. \cite{Heller,Dai}).

\section{Finite and noisy data}

So far we addressed the ideal case of a half-infinite and
noiseless sequence---also noiseless in the sense that the
expectation values are known precisely which requires infinite
statistics. It is, however, straight forward to analyze finite and
noisy data along the same lines. Let us begin with a finite
sequence $\avr{A(t)}$, $t\leq 2(N-1)$ and consider the $N\times N$
matrix $V_{kl}:=\avr{A(k+l-2)}$. As the rows of $V$ are a
truncation of the  vectors $v_{\tau}, \tau=0,\ldots, N-1$ we have
\be \dim\V\ \geq\ \text{rank} V,\ee with equality if $N\geq
\dim\cal V$. If the data are noisy or suffering a significant
statistical error, then  $V$ will typically be of full rank.
However, if an error estimate is available we may consider an
effective rank of $V$ by disregarding all singular values below a
certain noise threshold which is set by the estimated amount of
errors. More precisely, assume that $V$ is perturbed by some
$V_\epsilon$ (i.e., we actually observe $V'=V+V_\epsilon$) where
$||V_\epsilon||\leq\epsilon$. Then, by application of the singular
value inequality \cite{BhatiaIII64}, we get \be\label{eq:noise1}
\text{rank} V\geq\min\{k\;\big|\;s_{k+1}(V')\leq\epsilon\},\ee
where $s_l(V')$ is the $l$'th largest singular value of $V'$.

\begin{figure}[ttt]
\begin{center}
%\psfrag{title}{}
%\psfrag{xlabel}{\Large $\alpha$}
%\psfrag{ylabel}{\Large Benchmarks}
%\psfrag{legend1}{\large $\overline{F}$}
%\psfrag{legend2}{\large $\overline{F}_{infinite}$}
%\psfrag{legend3}{\large $\overline{F}_{finite}$}
\includegraphics[width=8cm]{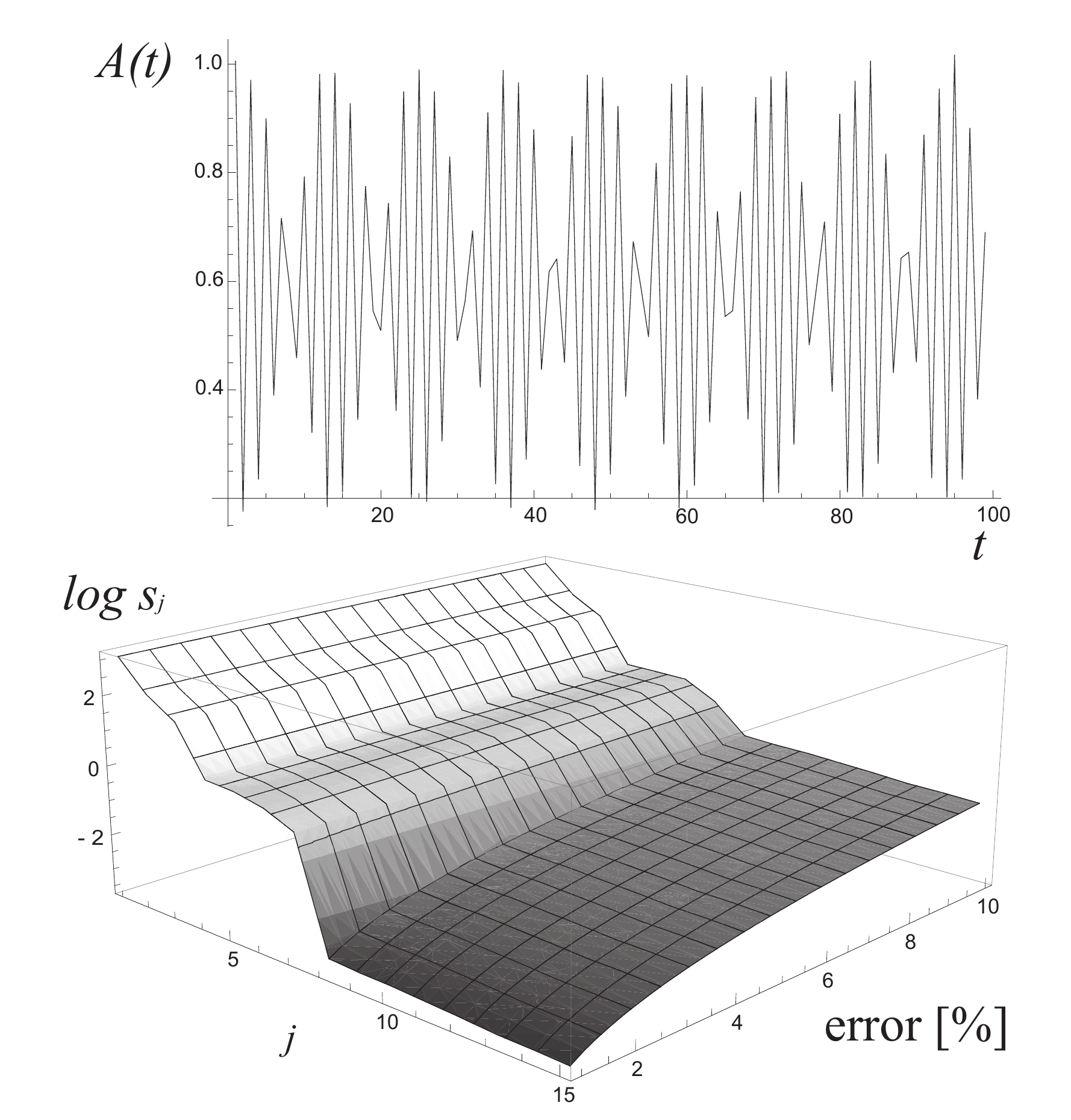}
%\epsfig{file=errorplotF.eps,angle=0,width=1\linewidth}
%\vspace*{-0.4cm}
\end{center}
\caption{\emph{top}: a finite sequence $\avr{A(t)}$, $t\leq 100$ obtained from a randomly chosen unitary dynamics of a spin-1 quantum system (d=3). As there are 3 conserved quantities (the eigenstates of the evolution) we have $\dim\V\leq 7$. \emph{bottom}: $\log$-plot of the 15 largest singular values $s_j$ of the corresponding $50\times 50$ matrix $V$ as a function of the standard deviation of added Gaussian noise (1\%-10\% of the signal). While for small noise the largest 7 singular values are clearly separated by a threshold, this washes out as the error increases---for too much noise the data could as well be explained by smaller $\dim\V$ as expressed quantitatively by Eq.(\ref{eq:noise1}). }
\label{Fig:noise}
\end{figure}

In Fig.\ref{Fig:noise} the behavior of these singular values is depicted graphically for an example of a unitary evolution $\tr{A U^t\rho U^{\dagger t}}$ with $d=3$ and randomly chosen $\rho, A, U$. As $U$ preserves its eigenstates and has (due to the random choice) no other conserved quantities, we have $\dim\V=7$ which is well reflected in the singular values of $V$ for small enough errors. In fact, in such unitary examples the dimension estimates appear to be surprisingly stable up to errors which make up a considerable fraction of the signal.

For the spectral analysis finite and noisy data seem to be more involved
to handle. It is not difficult to show \cite{Gantmacher} that
(\ref{eq:L}) is a rational function with exactly $\dim \V$ poles
and hence we are facing the problems of (i) giving a good (the
best) rational approximation (with a fixed number of poles) of a
function given by its Taylor coefficients and (ii) obtaining the
poles of a rational function given by its (noisy) Taylor
coefficients. The abstract solution to (i) is given by the
Adamyan-Arov-Krein Theorem \cite{Peller}. In practice (i.e.,
for finite data) a possible approach is the use of Pad\'e
approximants \cite{Pade}. For (ii), one can use for instance the
QD-algorithm of Rutishauer \cite{Herici}.

\section{Discussion}

In this letter, using tools from classical signal processing, we
have discussed how to determine the Hilbert space
dimension needed to explain  observed data of an evolving
quantum system. This introduces a new paradigm, beyond Bell
inequalities and the analysis of correlations, to obtain such
estimates with minimal assumptions in the model, in our case
homogeneity and Markovianity. In particular, the method can be
used for single systems, which answers a question posed in
\cite{Brunner} (for a different static approach, based on several input states, see the recent work \cite{Matthias}). We have also seen that we can use the method to quantify the non-Markovianity of an evolution via the effective dimension of the environments memory---complementing the static approach of \cite{snapshot}.
Finally, the analysis of spectral information revealed a dramatic
difference between the dimensions needed to give a quantum resp.
classical explanation of the data. Clearly, this does not outlaw a classical description (as the Garg-Leggett inequalities \cite{Leggett} do, albeit based on an extra arguable assumption) but it shows that quantum mechanics can be much more efficient in terms of number of degrees of freedom.
For the future, it would be nice to combine the two approaches, evolutions and
correlations, to obtain more information.

\

This work has been funded by Spanish grants I-MATH and
MTM2008-01366, by QUANTOP and the Ole Roemer grant of the Danish Natural Science Research Council (FNU). M.M.W. acknowledges discussion with M. Christandl and J.Appel.

\section*{Appendix A}

\begin{lemma}[Linear representation]\label{lem:linear}
For every bounded sequence $\avr{A(t)}\in\mathbb{C},\;t\in\mathbb{N}_0$ there are vectors $|L\rangle,|R\rangle$ of dimension $\dim\V$ and a respective matrix $M$ with $||M||_\infty\leq 1$ such that \be\label{eq:linear} \avr{A(t)}=\langle L|M^t|R\rangle.\ee
\end{lemma}
\proof{Consider $N\times N$ matrices $V_{kl}:=\avr{A(k+l-2)}$ and $V_{kl}':=\avr{A(k+l-1)}$ with $N$ large enough so that $\text{rank} V=\text{rank} V'=\dim\V<N$. Let $V=V_LV_R$ be a singular value decomposition where $V_R, V_L^T$ have $\dim\V$ rows. Then $|R\rangle:=V_R|1\rangle$, $\langle L|:=\langle 1|V_L$ and (using pseudo-inverses) $M:=V_L^{-1}V'V_R^{-1}$ satisfy Eq.(\ref{eq:linear}) obviously for $t\leq 2N-2$. For larger values of $t$, however, the same relation has to hold by linear dependence, i.e., the fact that by construction $\dim\V=\text{rank}V$.
boundedness of the sequence ($|\avr{A(t)}|\leq$ const.) implies that the spectral radius of $M$ is at most 1 and that eigenvalues of magnitude one have one-dimensional Jordan blocks (i.e. their geometric multiplicity equals the algebraic multiplicity). To see this first note that powers of a Jordan block $J(\lambda)$ are Toeplitz matrices with first row $J(\lambda)^t_{1k}=\lambda^{t-k+1}\left(\begin{array}{c}t \\ k-1  \\
\end{array}\right)$. Hence for $|\lambda|=1$ $\langle l|J(\lambda)^t|r\rangle$ is a polynomial in $t$ which remains only bounded for growing $t$ if it is zero.

Hence $M$ can only have Jordan blocks for $|\lambda|<1$. Let us finally show that  $M$ is therefore similar to a contraction: write $J(\lambda)=\lambda\1+N$ with $N$ a nilpotent matrix with ones on the first upper diagonal. Then $J(\lambda)$ is similar to $J':=\lambda\1+(1-|\lambda|)N$ and using the triangle inequality for the norm $||J'||_\infty\leq 1 $. As the representation of the sequence is preserved under similarity transformations of $M$ (and respective transformations of $R,L$) we can always w.l.o.g. chose $M$ to be a contraction.
}

\end{document}